\documentstyle[aps,prl]{revtex}
\draft
\begin{document}

\def\fecha{28 January 1998}

\twocolumn[\hsize\textwidth\columnwidth\hsize
\csname @twocolumnfalse\endcsname

\title{Quantum fields in anti-de Sitter wormholes}
\author{Carlos Barcel\'o$^{1,2}$ and Luis J. Garay$^2$}
\address{$^1$ Instituto de Astrof\'{\i}sica de
Andaluc\'{\i}a, CSIC,
Camino Bajo de Hu\'etor, 18080 Granada, Spain\\
$^2$ Instituto de Matem\'aticas y
F\'{\i}sica Fundamental, CSIC,
C/  Serrano 121, 28006 Madrid, Spain}
\date{\fecha}
\maketitle

\begin{abstract}

We study the effects of asymptotically anti-de Sitter wormholes
in low-energy field theory and give a general prescription for
obtaining the local effective interaction terms induced by them.
The choice of vacuum for the matter fields selects a basis of the
Hilbert space of anti-de Sitter wormholes whose elements can be
interpreted as states containing a given number of particles.
This interpretation is subject to the same kind of ambiguity in
the definition of particle as that arising from quantum field
theory in curved spacetime.

\end{abstract}

\pacs{04.60.Ds, 04.62.+v, 98.80 Hw \hfill {\it gr-qc/9703028}}
]

Among the possible kinds of fluctuations that might appear in
quantum gravity, wormholes describe processes that involve baby
universes branching off and joining onto asymptotically large
regions of spacetime. Wormholes have been extensively studied in
the literature \cite{colec1}. It has been claimed that, in the
dilute wormhole approximation, their effects in low-energy fields
can be taken into account through the introduction of local
effective interactions \cite{col,haw1}. As a consequence, they
would modify the coupling constants of any low-energy effective
physical theory. These effective interactions produced by
wormholes have been calculated for a variety of matter fields
coupled to gravity \cite{haw,lyo,dow,lafl,gar,luk}. In all these
calculations, the asymptotically large region of the wormhole has
been assumed to be flat. Therefore, the effective interactions
produced by such wormholes have been modelled by interacting
terms added to a free quantum field theory in flat spacetime.

In a previous paper \cite{nos}, we dealt with wormholes with
other types of asymptotic behaviour; in particular, we studied
the case of wormholes in asymptotically anti-de Sitter (adS)
spacetimes. This exhausts all possible maximally symmetric
asymptotic behaviours. Also, adS wormholes could be regarded as
excited states in the sense that the cosmological constant
$\Lambda$ could be interpreted as a non-vanishing asymptotic
energy of the matter field. We found an appropriate action for
these wormholes and showed that they give a non-vanishing
contribution to the path integral.

The calculation of the possible local effective interactions
produced by adS wormholes is of particular interest. We will see
that wormholes in adS spacetime do not contribute to the
cosmological constant nor to the gravitational coupling constant
directly. However, the presence of a cosmological constant will
certainly modify the form of the effective interactions produced
by wormholes, in comparison with the flat case, when these are
not spherically symmetric. Another interesting characteristic of
this model is that there naturally exist different possible
vacuum choices for the matter fields in the asymptotic regions,
as happens in quantum field theory in curved spacetime. There is
a relation between the definition of this vacuum and the choice
of basis in the Hilbert space of quantum wormholes that allows us
to interpret the basis elements as states of wormholes containing
certain number of particles.

The aim of this work is to model the effects of closed universes
that branch off or join onto an asymptotically adS spacetime by
means of an effective quantum field theory in an adS background.
We will consider processes that involve a single baby universe.
In the dilute wormhole approximation, the global effects will be
the result of these basic processes \cite{col,banks}. In order to
find these effects on the matter fields, and following previous
analyses in flat backgrounds \cite{haw,lyo,dow,lafl,gar,luk}, we
will calculate matrix elements of products of operators
$\Phi(x)$, each representing a generic matter field at a
different point, between an specific vacuum $|0\rangle$, and the
elements $|\Psi_\alpha \rangle$ of a basis of the Hilbert space
of adS wormholes \cite{haw}:
$$
\langle 0|\Phi(x_1)\cdots \Phi(x_r)|\Psi_\alpha\rangle .
$$
In flat spacetime, the state that is void of particles for
inertial observers defines a preferred vacuum for which its
associated propagator is asymptotically damped. In adS spacetime,
we can define a one-parameter family of maximally symmetric vacua
by analogy with the de Sitter case \cite{burges} (although there
are some subtleties in the definition of states in an adS
Lorentzian background because this spacetime does not posses a
well-posed Cauchy problem, we will work in the Euclidean sector
where this issue does not arise \cite{avis}). The propagator
associated with each vacuum of this family is damped when the
geodesic distance between the two points becomes large, but with
a different fall-off for each vacuum choice \cite{allen1}. This
family of vacua is the analogue in adS spacetime of the vacua
defined by uniformly accelerated reference frames in the flat
case. The state $|0\rangle$ is any fixed vacuum of this family.

The wormhole states $|\Psi \rangle$ will be represented by wave
functions $\Psi[g_{ij}',\Phi']$ that are solutions of the
Wheeler-deWitt equation and satisfy the wormhole boundary
conditions \cite{page}. The arguments of these wave functions are
the three-geometry and the matter content of the system under
consideration. We will restrict our study to spacetime geometries
that can be foliated in nearly spherically symmetric
three-surfaces, which will be described by a scale factor and
small perturbations around the homogeneous configuration, and
bosonic massless fields as matter content. In an appropriate
gauge, we can just keep as configuration variables the scale
factor and the different coefficients that represent the true
degrees of freedom of the matter fields in an expansion in terms
of hyperspherical harmonics on the three-sphere \cite{lifs,hall}.
For conformally coupled matter fields, such as a conformal scalar
field or an electromagnetic field, the scale factor decouples
from the matter degrees of freedom in the Wheeler-deWitt
equation. The part of the wave function that contains the scale
factor is asymptotically damped and indicates the existence of an
asymptotically adS region. On the other hand, each mode $n$ in
the harmonic expansion of the matter content can be described as
a quantum harmonic oscillator with a frequency that only depends
on the specific mode $n$. As we will see, each quantum of energy
of these oscillators associated with the mode $n$ can be
interpreted as a particle of mode $n$ in the conformal vacuum,
that is, in the vacuum conformally related to the natural vacuum
in flat spacetime.

When the matter field is coupled in a non-conformal way to
gravity, as happens with gravitons \cite{gris}, we cannot carry
out the above separation of geometrical and matter degrees of
freedom. The different matter harmonic oscillators would have
frequencies that depend on the scale factor and therefore we
could not interpret them as particles. The creation
of particles generated by the expansion of these
systems do not allow us to interpret the different wormhole
states as containing certain number of particles.

Finally, an important restriction on wormhole states is due to
the linearisation instability of the perturbations to the
rotationally invariant configurations \cite{lafl,mon}. In other
words, the invariance under the isotropy group $SO(4)$ restricts
the Hilbert space of wormhole states to those that contain only
rotationally invariant matter states. This means that, for the
inhomogeneous modes, single-particle wormhole states are not
allowed.

We are interested in processes in which there is creation or
annihilation of particles associated with a specific vacuum in
two different points of a spacetime region where a wormhole end
is inserted. The matrix element $\langle 0|\Phi(x_1)
\Phi(x_2)|\Psi_\alpha\rangle $ is then given, following Hawking
\cite{haw}, by a path integral over geometries and matter fields
that induce a metric $g_{ij}'$ and a value $\Phi'$ for the matter
field in a cross section of the wormhole and that satisfy some
asymptotic requirements that guarantee the asymptotically adS
behaviour and that select a vacuum for the matter field in the
semiclassical approximation. Then, we will integrate over all
possible configurations $(g_{ij}'$, $\Phi')$ weighted with the
wormhole wave function. Thus, the matrix element has the form
\begin{eqnarray}
&&\langle 0|\Phi(x_1) \Phi(x_2)|\Psi_\alpha\rangle = \nonumber\\
&&\int {\cal D}\Phi' {\cal D}g_{ij}' \Psi_\alpha [g_{ij}',\Phi']
\int {\cal D}\Phi {\cal D} g_{\mu \nu} \Phi(x_1) \Phi(x_2)
e^{-I[g_{\mu \nu},\Phi]},\nonumber
\end{eqnarray}
plus the appropriate gauge fixing condition and Fadeev-Popov
determinant. We will evaluate this path integral semiclassically.
The action $I[g_{\mu \nu},\Phi]$ contains a surface term that
renders it finite for classical solutions \cite{nos}, a necessary
requirement for the semiclassical approximation to be meaningful.
The saddle point solution for the gravitational part must be an
asymptotically adS spacetime. As far as the low-energy regime is
concerned, it can be taken to be pure adS spacetime outside a
three-sphere in which the wave function takes its arguments.

The position of the spacetime points $x_1$, $x_2$ can only be
specified modulo the isometries of Euclidean adS spacetime,
$SO(4,1)$. This group is isomorphic to the semi-direct product of
adS translations and the isotropy group $SO(4)$. If $\Phi(x)$ is
a saddle point solution for the matter field, then,
$\Phi_{x'h}(x)$, the transform of $\Phi(x)$ under the isometry
$(x'h)^{-1}$, is also a solution, $x'$ being an element of the
group of translations in adS spacetime and $h\in SO(4)$. As a
consequence, one has to average over $SO(4,1)$. The invariant
Haar measure in this group is equivalent to integrating first
over the group $SO(4)$ and, then, over the coset space, i.e. over
adS spacetime \cite{bar}. We can interpret this integral as an
average over the orientations $h$ of the wormhole and the
positions $x'$ in which it can be inserted. In flat spacetime,
similar results can be obtained, if we use its own group of
isometries, the Euclidean group in four dimensions $E_4$
\cite{haw}.

In an appropriate gauge, we can expand the massless bosonic field
under consideration in terms of hyperspherical harmonics in a way
that only the true matter field degrees of freedom are present.
Let
$$
\Phi_{x'h}(x)=\sum_{n,\sigma_n} c_n[\mu(x,x')] M_h^{n,\sigma_n}
{\cal H}_{n,\sigma_n}(\widehat{x-x'})
$$
be such expansion, where $n$ is the mode of the hyperspherical
harmonic ${\cal H}_{n,\sigma_n}(\widehat{x-x'})$, the hat denotes
unit vectors, $c_n[\mu(x,x')]$ are functions of the geodesic
distance $\mu(x,x')$ between $x$ and $x'$ and $\sigma_n$ is an
index that covers the degeneration space of the harmonic mode
$n$. This space carries an irreducible representation of the
group $SO(4)$ \cite{bar}. If we use a tensor form for this
representation, $M_h^{n,\sigma_n}$ will denote the components
$\sigma_n$ of a tensor that belongs to the representation
associated with the mode $n$ and rotated by an element $h$ of
$SO(4)$. Taking into account that these irreducible tensor
representations of $SO(4)$ are characterised by the different
kinds of symmetries under permutations of their tensor indices
$\sigma_n$ \cite{lifs} and that the propagator of the matter
field in adS spacetime $G(x,x')$ and the saddle point
$\Phi_{x'h}(x)$ satisfy the same equation, we can easily conclude
that this saddle point can be expressed for each
harmonic mode $n$ in terms of the propagator $G$ alone.
Explicitly, $\Phi_{x'h}^{n}(x)={\cal M}_h^n \cdot \Theta_n'
G(x,x')$, where now $ {\cal M}_h^n$ is a constant tensor that
contains all the dependence on the arguments of the wave function
and the rotation group, and the dot represents a contraction of
all primed tensor indices. The operator $\Theta_n$ can be
constructed recursively by symmetrising $\nabla\Theta_{n-1}$ and
substracting all its traces. The operator $\Theta_{n_0}$ for the
lowest mode is given by the smallest irreducible representation
of each matter field and is therefore specific to each matter
content: it is the identity for the scalar field, the
antisymmetric derivative in the electromagnetic case and, for
gravitons, it has the same derivative structure as the Weyl
tensor $C^{\mu\nu\rho\sigma}$. Furthermore, for fields with spin,
like photons and gravitons, we have to deal with both helicities,
positive and negative, separately. This amounts to introducing
operators $\Theta_{n\pm}$ which are the self-dual and
anti-self-dual parts of $\Theta_{n}$, i.e.
$\Theta_{n\pm}=\Theta_{n}\pm{^*\Theta}_{n}$, the star denoting
the Hodge dual. For fermions or massive fields, the
representation theory provides analogous results, although the
explicit expressions are slightly more involved.

In view of the simple form of the saddle points, it is easily
seen that the wormhole matrix element contains the product
$[{\cal M}_h^n \cdot \Theta_n' G(x_1,x')] [{\cal M}_h^n \cdot
\Theta_n' G(x_2,x')]$. Using the property \cite{bar} that, for
any irreducible representation $R(h)$ of the group $SO(4)$,
$\int_{SO(4)} dh R(h)\otimes R(h)=1\otimes 1$, the average over
rotations yields $({\cal M}^n \cdot {\cal M}^n) [\Theta_n'
G(x_1,x') \cdot \Theta_n' G(x_2,x')]$ and the matrix element
becomes
\begin{eqnarray}
&&\langle 0|\Phi(x_1) \Phi(x_2)|\Psi_\alpha\rangle= \nonumber\\
&&K(0,\Psi_{\alpha})\int d^4x' \sqrt{g(x')} \Theta_n' G(x_1,x')
\cdot \Theta_n' G(x_2,x')\nonumber
\end{eqnarray}
where $K(0,\Psi_{\alpha})$ is a constant that depends on the
choice of vacuum and on the state $|\Psi_{\alpha}\rangle$.

For conformal matter fields and choosing the conformal vacuum,
the path integral that appears in the definition of this
two-point matrix element gives separate contributions for the
scale factor and the matter harmonic modes. The only wave
function that gives a nonvanishing result for
$K(0,\Psi_{\alpha})$ is that which contains only two quanta in
the mode $n$, apart from the vacuum itself \cite{haw,dow}. A
similar analysis, applied to matrix elements of more than two
points, allows us to interpret each state of the basis of the
Hilbert space of wormholes characterised by the number of energy
quanta of the matter harmonic oscillator in the mode $n$, as a
quantum wormhole that contains a given number of particles in
that mode associated with the conformal vacuum. For other vacuum
choices, the path integral does not separate but still, analogous
interpretations for different bases of the wormhole Hilbert space
can be reached. Therefore, for each vacuum choice, there exists
an orthonormal basis in the Hilbert space of wormholes such that
its elements can be labelled by the number of particles that they
contain. Thus, the ambiguity in the choice of vacuum which is
present in quantum field theory in curved spacetimes also shows
up in wormhole physics. When the matter content is not
conformally coupled, as in the graviton case, we cannot label the
wormhole states by their particle content because there exists
creation of particles owing to the expansion of the asymptotic
region. The states are global states particle-background.

Let us now find an interaction Lagrangian that, via the formula
$$
\langle 0|\Phi(x_1)\Phi(x_2) \int d^4x' \sqrt{g(x')} {\cal L}_I
\left[\Phi (x')\right]|0\rangle,
$$
where, here, $|0\rangle$ represents a matter field vacuum in adS
background, reproduces the matrix element above, up to a constant
factor. It is straightforward to see that the required
interaction Lagrangian is
$$
{\cal L}_I=(\Theta\Phi)^2.
$$
Indeed, the linearity of the operator $\Theta$ and Wick's theorem
allow us to write the two-point function above in the form
$$
\int d^4x' \sqrt{g(x')} \Theta'G(x_1,x') \cdot \Theta'G(x_2,x').
$$

Let us now apply these results to three separate kinds of matter
fields, namely, a conformal scalar field, and an electromagnetic
field, which are conformally coupled to gravity, and gravitons,
which are not because of the presence of a cosmological constant.
For the lowest mode, this Lagrangian reads ${\cal L}_I=\phi^2$
for the scalar field, ${\cal L}_I=F_{\mu\nu}F^{\mu\nu}$ for
photons and ${\cal L}_I=C_{\mu\nu\rho\sigma}C^{\mu\nu\rho\sigma}$
for gravitons, which coincide with the results obtained for
asymptotically flat wormholes \cite{haw,lyo,dow,lafl}. In the
case of photons and gravitons, positive and negative helicities
provide the same interaction Lagrangian, as expected, because the
cross product $\Theta_n\Phi \cdot {^{*}\Theta}_n\Phi$ is a
topological invariant, as can be checked by direct calculation.
If we call $n_0$ to this lowest mode for each matter content
($n_0=1,2,3$ for scalars, photons and gravitons, respectively),
the operator $\Theta_{n_0+1}$ for the next mode is, following the
prescription described above, $\nabla\Theta_{n_0}$, where the
required symmetrisation is automatically implemented. Therefore,
the effective interaction Lagrangian for this mode $n_0+1$, as
well as that for the lowest one $n_0$, does not depend on the
cosmological constant and has the same form as that obtained in
flat spacetime. For the next higher mode, $\Theta_{n_0+2}$
contains a symmetrised and traceless combination of products of
two derivatives. In particular, it will contain a term of the
form $g^{\mu\nu}\nabla^2\Theta_{n_0}$, which is proportional to
the cosmological constant by virtue of the equation for the
propagator in adS spacetime. For instance, the effective
interaction Lagrangian coming from the $n=3$ inhomogeneous mode
of a scalar field is of the form $(\nabla^{\mu}\nabla^{\nu}\phi
-\frac{1}{6}\Lambda g^{\mu\nu}\phi)^2$. We then see that the
interaction Lagrangians for higher modes in adS spacetime and
those in the flat case differ in terms that are proportional to
powers of the cosmological constant. It is also worth noting that
adS wormholes do not induce any direct modification to the
cosmological term nor to Newton's constant.

Let us conclude with a brief summary. We have given a general
prescription for finding the low-energy effective interaction
Lagrangians induced by wormholes in a maximally-symmetric
asymptotically large region of spacetime. We have seen that the
two lowest modes in the harmonic expansion of each matter field
give interaction Lagrangians of the same form as those obtained
in asymptotically flat spacetimes, while the higher inhomogeneous
modes give Lagrangians that contain an explicit dependence on the
cosmological constant.

We are very grateful to G.A. Mena Marug\'an, M. Moles and P.F.
Gonz\'alez-D\'{\i}az for helpful discussions and suggestions.
C.B. was supported by a Spanish Ministry of Education and Culture
(MEC) grant. C.B. is also grateful to J.B. Hartle and the
Institute for Theoretical Physics (UCSB), where part of this work
was done, for warm hospitality. This research was supported in
part by the National Science Foundation under Grant No.
PHY994--07194. L.J.G. was supported by funds provided by DGICYT and
MEC (Spain) under Contract Adjunct to the Project No. PB94--0107.


\begin{thebibliography}{99}

\bibitem{colec1} A. Strominger, in {\it Particles, strings and
supernovae}, edited by A. Jevicki and C.I. Tan (World Scientific,
Singapore, 1989); S.W. Hawking, Mod. Phys. Lett. {\bf A5}, 145
(1990); {\it ibid.} {\bf A5}, 453 (1990); L.J. Garay, {\it
Agujeros de gusano en cosmolog\'{\i}a cu\'antica}, Ph. D. Thesis
(Universidad Aut\'onoma de Madrid, 1992) and references therein.

\bibitem{col} S. Coleman, Nucl. Phys. {\bf B307}, 867 (1988).

\bibitem{haw1} S.W. Hawking, Nucl. Phys. {\bf B335}, 155 (1990).

\bibitem{haw} S.W. Hawking, Phys. Rev. D {\bf 37}, 904 (1988).

\bibitem{lyo} A. Lyons, Nucl. Phys. {\bf B324}, 253 (1989).

\bibitem{dow} H.F. Dowker, Nucl. Phys. {\bf B331}, 194 (1990).

\bibitem{lafl} H.F. Dowker and R. Laflamme, Nucl. Phys. {\bf
B366}, 209 (1991).

\bibitem{gar} L.J. Garay and J. Garc\'{\i}a-Bellido, Nucl. Phys.
{\bf B400}, 416 (1993).

\bibitem{luk} A. Lukas, Nucl. Phys. {\bf B442}, 533 (1995).

\bibitem{nos} C. Barcel\'o, L.J. Garay, P.F. Gonz\'alez-D\'{\i}az
and G.A. Mena Marug\'an, Phys. Rev. D {\bf 53}, 3162 (1996).

\bibitem{banks} L. Klebanov, L. Susskind, T. Banks, Nucl. Phys.
{\bf B317}, 665 (1989).

\bibitem{burges} C.J.C. Burges, Nucl. Phys. {\bf B247}, 533
(1984). B. Allen, Phys. Rev. D {\bf 32}, 3136 (1985).

\bibitem{avis} S.J. Avis, C.J. Isham and D. Storey, Phys. Rev.
D {\bf 18}, 3565 (1978).

\bibitem{allen1} B. Allen and T. Jacobson, Commun. Math. Phys.
{\bf 103}, 669 (1986).

\bibitem{page} S.W. Hawking and D.N. Page, Phys. Rev. D {\bf
42} 2655 (1990).

\bibitem{lifs} E.M. Lifshitz, J. Phys. {\bf 10}, 116 (1946); E.M.
Lifshitz and I. Khalatnikov, Adv. Phys. {\bf 12}, 185 (1963).

\bibitem{hall} J.J. Halliwell and S.W. Hawking, Phys. Rev. D {\bf
31}, 1777 (1985).

\bibitem{gris} L.P. Grishchuk, Zh. Eksp. Teor. Fiz. {\bf 67}, 825
(1974); Sov. Phys. JETP {\bf 40}, 409 (1975). S. Wada, Nucl.
Phys. {\bf B276}, 729 (1986).

\bibitem{mon} V. Moncrief, Phys. Rev. D {\bf 18}, 983 (1978).

\bibitem{bar} A. Barut and R. Raczka, {\it Theory of Group
Representations and Applications} (World Scientific, Singapore,
1986).

\end{thebibliography}
\end{document}